\documentclass[pra,amsmath,aps,notitlepage,amssymb,showpacs,twocolumn,superscriptaddress]{revtex4-1}
\usepackage{amsthm}
\usepackage{bm}
\usepackage{graphicx}
\usepackage{color}

\newtheorem*{theorem*}{Theorem}
\newtheorem{Def}{Definition}
\newtheorem{Thm}[Def]{Theorem}
\newtheorem{Lem}[Def]{Lemma} 
\newtheorem{Ex}[Def]{Example}
\newtheorem{Cor}[Def]{Corollary}
\newcommand{\bdf}{\begin{Def}}
\newcommand{\edf}{\end{Def}}
\newcommand{\bex}{\begin{Ex}}
\newcommand{\eex}{\end{Ex}}
\newcommand{\bthm}{\begin{Thm}}
\newcommand{\ethm}{\end{Thm}}
\newcommand{\blm}{\begin{Lem}}
\newcommand{\elm}{\end{Lem}}
\newcommand{\bcor}{\begin{Cor}}
\newcommand{\ecor}{\end{Cor}}

\newcommand{\ot}{\otimes}

\newcommand*{\cH}{\mathcal{H}}

\newcommand*{\cM}{\mathcal{M}}

\newcommand*{\cO}{\mathcal{O}}

\newcommand*{\cS}{\mathcal{S}}
\newcommand*{\cU}{\mathcal{U}}

\newcommand*{\tr}{\mathrm{Tr}}

\newcommand{\argmax}{\mathop{\rm arg~max}\limits}
\newcommand{\argmin}{\mathop{\rm arg~min}\limits}
\newcommand{\trho}{{\tilde \rho}}
\newcommand{\tsigma}{{\tilde\sigma}}

\newcommand*{\id}{\rm id}
\def\idty{{\leavevmode\rm 1\mkern -5.4mu I}}

\usepackage{color}
\definecolor{myred}{rgb}{1,0,0}

\definecolor{myblue}{rgb}{0,0,0.8}

\definecolor{myyellow}{rgb}{0.9,0.8,0}

\definecolor{mygreen}{rgb}{0,0.6,0}

\definecolor{myorange}{rgb}{0.6,0.6,0}

\definecolor{mycerul}{rgb}{0,0.6,1}

\begin{document}

\title{Information-theoretical analysis of topological entanglement entropy and multipartite correlations}

\author{Kohtaro Kato}
\affiliation{Department of Physics, Graduate School of Science, The University of Tokyo, Tokyo, Japan}

\author{Fabian Furrer}
\affiliation{NTT Basic Research Laboratories, NTT Corporation, 3-1 Morinosato Wakamiya, Atsugi, Kanagawa, 243-0198, Japan}
\affiliation{Department of Physics, Graduate School of Science, The University of Tokyo, Tokyo, Japan}

\author{Mio Murao}
\affiliation{Department of Physics, Graduate School of Science,
The University of Tokyo, Tokyo, Japan}
\affiliation{Institute for Nano Quantum Information Electronics,
The University of Tokyo, Tokyo, Japan}

\date{\today}

\begin{abstract}
A special feature of the ground state in a topologically ordered phase is the existence of  large scale correlations depending only on the topology of the regions. These correlations can be detected by the topological entanglement entropy or by a measure called irreducible correlation. We show that these two measures coincide for states obeying an area law and having zero-correlation length. Moreover, we provide an operational meaning for these measures by proving its equivalence to the optimal rate of a particular class of secret sharing protocols. This establishes an information-theoretical approach to multipartite correlations in topologically ordered systems.  
\end{abstract}
\maketitle

\section{Introduction}
Topologically ordered phase is an exotic quantum phase that cannot be explained by conventional models based on {\it local} order parameters and symmetry-breaking~\cite{PhysRevB.40.7387,PhysRevB.41.9377}.  
Since topologically ordered phases are robust against local perturbations, they are promising candidates for performing topologically-protected quantum computation~\cite{Freedman2002,Kitaev2003a}.
Characterizing the global properties of the ground state in topologically ordered phases is thus not only an important problem in condensed matter physics, but also in quantum information science. 

\begin{figure}
\begin{center}
\vspace{-5mm}
\includegraphics[width=7cm]{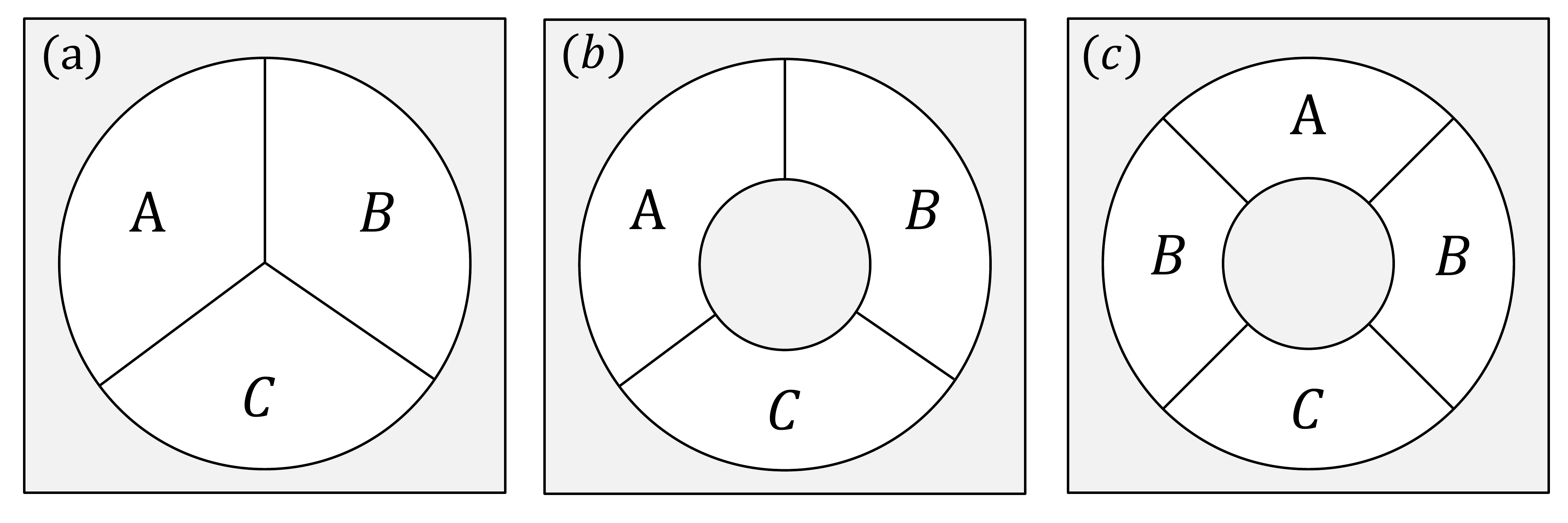}
\vspace{-5mm}
\end{center}
\caption{
The illustration shows the regions $A,B,C$ used for the calculation of the TEE and the irreducible correlation. Region $(a)$ and $(b)$ correspond to the Kitaev-Preskill type version~\cite{PhysRevLett.96.110404} and $(c)$ to the Levin-Wen type version~\cite{PhysRevLett.96.110405} of the TEE. Due to the difference of the topology of the regions, the value of the TEE for $(a)$ is half of the ones for $(b)$ and $(c)$.}
\label{region}
\vspace{-5mm}
\end{figure}

A possible measure to detect such topological correlations is the topological entanglement entropy (TEE)~\cite{PhysRevLett.96.110404,PhysRevLett.96.110405}, which also appears as the universal constant term in the area law~\cite{Hamma200522, PhysRevLett.96.110404}.
The definition of the TEE is based on the idea that topological correlations reduce the entropy of ring or circle-like regions (see Fig.~\ref{region}), compared to what is expected if only short-range correlations are present~\cite{PhysRevLett.96.110405}.  
More precisely, the TEE quantifies the tripartite correlations in region $ABC$ that are not contained in any bipartite region $AB$, $BC$, or $CA$. Quantitatively, this is achieved by subtracting the contributions of local correlations using a Venn-diagram calculation, which is known as the interaction information in classical information theory~\cite{coinfo1954}. 
The interaction information was proposed as a measure of ``genuine" tripartite correlations that only detects correlations shared by all three parties, but not by only two.
However, the information theoretical meaning of the interaction information is not clear, since it lacks basic properties such as, e.g., positivity (see, e.g.,~\cite{Watanabe:1960, krippendorff09, 2010Entrp..12...63L}). Further problems arise in the quantum case, where, for instance, the quantity is always zero if the three parties share a pure state.

An alternative measure for ``genuine"  tripartite or more generally, $k$-partite  correlations in classical information theory is known as  the {\it $k$th-order effect}~\cite{amari930911}. 
The definition of the $k$th-order effect employs the maximum entropy method~\cite{PhysRev.106.620,PhysRev.108.171} to estimate the total entropy, which provides a classification of multipartite correlations in terms of Gibbs states corresponding to $k$-local Hamiltonians. 
The quantum generalization of the $k$th-order effect is called the {\it $k$th-order (or $k$-body) irreducible correlation}~\cite{PhysRevLett.89.207901,PhysRevLett.101.180505,PhysRevA.80.022113}.

 The 3rd-order irreducible correlation applied to tripartite scenarios as shown in Fig~\ref{region} has recently been proposed as an alternative way to measure topological correlations~\cite{2014arXiv1402.4245L,2014arXiv1406.5046C}. It is simply given as the maximum entropy on $ABC$ with consistent bipartite reduced states on $AB$,$BC$, $AC$ minus the actual entropy of $ABC$. 
It has been conjectured that the 3rd-order irreducible correlation and the TEE coincide in the thermodynamic limit for ground states of gapped systems, i.e., when the effects of local correlations are completely negligible~\cite{2014arXiv1406.5046C}. While therein the authors only show that the irreducible correlation is always smaller than the TEE for regions as in Fig.~\ref{region}$(c)$, numerical evidence of this conjecture has been provided for the toric code model~\cite{Kitaev2003a} in Ref.~\cite{2014arXiv1402.4245L}. 

In this paper, we partly resolve this conjecture and show that if the ground state obeys an area law and has exactly vanishing correlation lengths, the TEE and the 3rd-order irreducible correlation are equivalent.  This  condition holds for a wide class of spin-lattice models, which describe non-chiral topological ordered phases~\cite{Kitaev2003a, PhysRevB.71.045110}. 
In general, calculating the values of multipartite correlation measures is a computationally hard problem. We overcome this challenge and show equivalence by explicitly constructing the maximum entropy state on $ABC$ that is consistent with all bipartite reduced density matrices (RDMs) of the ground state. This solves an instance of a quantum marginal problem~\cite{2004quant.ph..9113K,2006quant.ph..4166L}, which is in general hard, especially, if RDMs have overlap. In our special case the difficulty can be overcome by using properties of {\it quantum Markov states} (QMS)~\cite{ssaiff2004}.

We further show that under the same assumptions the irreducible correlation is equal to the optimal asymptotic rate of a secret sharing protocol as suggested in~\cite{2007quant.ph..1029Z,PhysRevLett.101.180505}.
This establishes an operational interpretation of the TEE as the number of bits that can be hidden in a global region $ABC$ (see Fig.~\ref{region}) from parties that have only access to partial regions such as, e.g., $AB$.  
This result quantitatively connects the TEE with the characteristic feature of topologically ordered states that information contained in local regions are insufficient to determine global properties.

The paper is organized as follows. In Section~\ref{sec:irr}, we define the irreducible correlation of a multipartite state and discuss its properties. 
In Section~\ref{sec:topo}, we prove the equivalence of the TEE and the irreducible correlation.  In Section~\ref{sec:ss}, we show the equivalence of the irreducible correlation and the maximum rate of a secret sharing protocol. Section~\ref{sec:Approx} is devoted to a discussion of the case of almost vanishing correlations. Our conclusions are presented in Section~\ref{sec:Conclusion}.

\section{The irreducible correlation}~\label{sec:irr}

Let $\rho$ be an $n$-partite state in a state space $\cS(\cH^n)$, where $\cH^n=\cH_1\ot\cdots\ot\cH_n$ and each Hilbert space $\cH_i$ is finite-dimensional. The RDM on the subsystem corresponding to a subset $A$ of $[n]\equiv\{1,...,n\}$ is denoted by $\rho_A$. 
We then define a closed convex set $R_\rho^k$ of $n$-partite states, where all their $k$-RDMs are identical to $\rho$ as
\begin{equation}\label{defR}
R_\rho^k\equiv\left\{\sigma\in\cS(\cH^{n}) \mid \forall S_k\subset[n], \;|S_k|=k :\,\sigma_{S_k}=\rho_{S_k}\right\}\,.
\end{equation}

For any $\rho$ and $1\leq k\leq n$, we define the $k$th-maximum entropy state ${\tilde \rho}^{(k)}\in\cS(\cH)$ by the state in $R_\rho^k$ which maximizes the von Neumann entropy, i.e.,
\begin{equation}\label{maxent}
\trho^{(k)}\equiv\argmax_{\sigma\in R_\rho^k}S(\sigma)\,, 
\end{equation}
where $S(\rho)=-\tr \rho\log\rho$.
According to Jaynes\rq{}s maximal entropy principle~\cite{PhysRev.106.620,PhysRev.108.171}, $\trho^k$ is the most \lq\lq{}unbiased\rq\rq{} inference if all of the $k$-RDMs of $\rho$ are known.

The $k$th-maximum entropy state can be characterized by quantum Gibbs families. Let $Q_k$ be the set of Gibbs states $e^H/\tr(e^H)$ corresponding to $k$-local Hamiltonians $H$. 
We consider the reverse information-closure $\bar{Q}_k=\{\sigma\in S(\cH^n) \vert \inf_{\sigma\rq{}\in Q_k} S(\sigma\Vert\sigma\rq{})=0\}$ of $Q_k$, where $S(\rho||\sigma)=\tr\rho\log\rho-\tr\rho\log\sigma$ is the quantum relative entropy.
Then $\trho^{(k)}$ satisfies~\cite{weismaxent2015}
\begin{equation}\label{infproj}
\trho^{(k)}=\argmin_{\sigma\in \bar{Q}_k}S(\rho||\sigma)\,.
\end{equation}
For any state $\rho\in\cS(\cH^n)$ and $1\leq k\leq n$ follows that $\trho^{(k)}$ is uniquely determined and the Pythagorean theorem~\cite{2010.5671W} 
\begin{equation}\label{Pytha}
S(\rho||\sigma)=S(\rho||\trho^{(k)})+S(\trho^{(k)}\Vert\sigma)\,
\end{equation}
holds.
Since the completely mixed state $\idty_n/d^n$ is in $\bar Q_k$ for all $k$, the Pythagorean theorem implies that 
\begin{equation}
S(\rho||\trho^{(k)})=S(\trho^{(k)})-S(\rho)\,.
\end{equation}

We define $D^{(k)}(\rho)$ as the distance of the state $\rho$ from the set $\bar Q_k$, that is,
\begin{align}
D^{(k)}(\rho) & \equiv \min_{\sigma\in \bar{Q}_k}S(\rho||\sigma) \\& =S(\rho||\trho^{(k)})=S(\trho^{(k)})-S(\rho)\,.
\end{align}
Since ${\bar Q}_{k-1}\subset {\bar Q}_k$, it holds that $D^{(k)}(\rho)$ is monotonically decreasing in $k$. 
From the Pinsker inequality, it is clear that $D^{(k)}(\rho)$ measures how well the state $\rho$ is approximated by the maximum entropy principle, i.e., 
 $D^{(k)}(\rho)\leq\epsilon$ implies that $\Vert\rho-\trho^{(k)}\Vert_1\leq 2\sqrt{\epsilon}$. It further holds that $D^{(k)}(\rho)=0$, if and only if $\rho=\trho^{(k)}\in{\bar Q}_k$.

\begin{figure}
\begin{center}
\includegraphics[width=6cm]{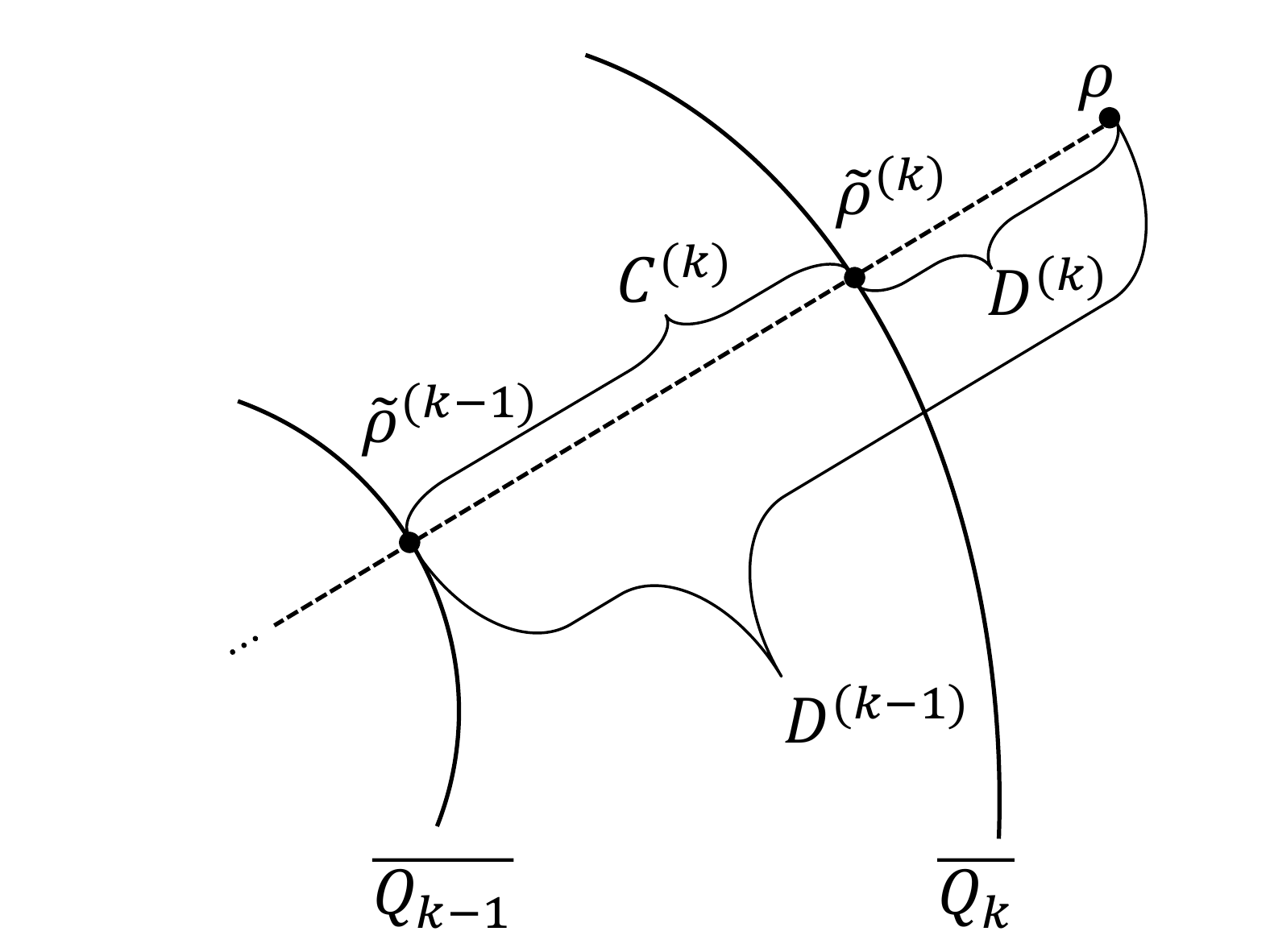}
\end{center}\vspace{-5mm}
\caption{A geometrical illustration of the functions $D^{(k)}$ and $C^{(k)}$. $D^{(k)}(\rho)$ is the distance from the set of  $k$-correlated states and $C^{(k)}$ is the difference between crossing points in ${\bar Q}_k$ and ${\bar Q_{k-1}}$ measured by the quantum relative entropy.}
\label{fig:dist}
\end{figure}

The $k$-th order irreducible correlation~\cite{PhysRevLett.101.180505} is defined as
\begin{align}\label{eq:irrdef}
C^{(k)}(\rho)&\equiv D^{(k-1)}(\rho)-D^{(k)}(\rho)\\
&=S(\trho^{(k)}\Vert\trho^{(k-1)})\\
&=S({\tilde \rho}^{(k-1)})-S({\tilde \rho}^{(k)})\,.
\end{align}
The second equation follows from the Pythagorean theorem Eq.~\eqref{Pytha} and the fact that $\trho^{(k)}$ has the same $(k-1)$-RDMs as $\trho^{(k-1)}$. 
A geometric picture of the relations between $D^{(k)}$ and $C^{(k)}$ is given in Fig.~\ref{fig:dist}. The classical analogue of the above discussion has been given in Ref.~\cite{amari930911}.

$C^{(k)}(\rho)$ measures the correlation that is contained in all the $k$-RDMs, but not in the $(k-1)$-RDMs. 
The irreducible correlation is a non-negative function invariant under local unitary operations on each single site, but lacks a non-increasing nature under general local operations~\cite{PhysRevA.80.022113,PhysRevE.85.046209}.
A possible modification of the irreducible correlation that overcomes this problem is proposed in Ref.~\cite{PhysRevE.85.046209}.
The irreducible correlation is continuous in the classical case~\cite{Barndorff-Nielsen197804}, but it can be discontinuous for quantum states~\cite{weisdiscont2012}.
A relation between the discontinuity of the irreducible correlation and quantum phase transitions has been discussed in Ref.~\cite{2014arXiv1406.5046C}. 

We show now that $C^{(k)}(\rho)$ is also additive $C^{(k)}(\rho\ot\sigma)=C^{(k)}(\rho)+C^{(k)}(\sigma)$. It is clear that $\trho^{(k)}\ot\tsigma^{(k)}$ is included in $R_{\rho\ot\sigma}^k$. Let us consider $\rho\in\cS(\cH_1)$ and $\sigma\in\cS(\cH_2)$. For any state $\omega_{12}\in R_{\rho\ot\sigma}^k$, it holds that 
\begin{align}
S(\omega_{12})&\leq S(\omega_1) +S(\omega_2)\\&\leq S(\trho^{(k)})+S(\tsigma^{(k)})\\&=S(\trho^{(k)}\ot\tsigma^{(k)}),
\end{align}
where we used the subadditivity of the von Neumann entropy and the fact that  $\omega_{12}\in R_{\rho\ot\sigma}^k$ implies $\omega_1\in R_\rho^k$ and $\omega_2\in R_\omega^k$. 
Therefore, $\trho^{(k)}\ot\tsigma^{(k)}$ is the $k$th-maximum entropy state corresponding to $\rho\ot\sigma$. By definition, this implies that the irreducible correlation is additive. 

The total correlation~\cite{Watanabe:1960, PhysRevLett.104.080501} of an  $n$-partite state $\rho$ is given by 
\begin{equation}\label{eq:todef}
C^T(\rho)\equiv\sum_{i=1}^nS(\rho_i)-S(\rho)\,.
\end{equation}
This function is considered as one of the generalizations of the mutual information for multipartite states. 
From the definition of $C^{(k)}(\rho)$ in Eq.~\eqref{eq:irrdef} and by using the fact that $\trho^{(1)}=\rho_1\ot\cdots\ot\rho_n$ and $\trho^{(n)}=\rho$, we can decompose the total correlation into the sum of $k$-th order irreducible correlations, i.e.,
\begin{equation}\label{eq:toc}
C^T(\rho)=D^{(1)}(\rho)=\sum_{k=2}^nC^{(k)}(\rho)\,.
\end{equation}
\section{Equivalence of  TEE and the irreducible correlation} \label{sec:topo}
Let us consider the reduced state of the ground state of a gapped spin lattice system on circle or ring-like regions $ABC$ given in Fig.~\ref{region}. 
We then define the TEE by 
\begin{align}\label{deftopo}
\gamma&\equiv S_\rho(AB)+S_\rho(BC)+S_\rho(CA)\nonumber\\
&\quad-S_\rho(A)-S_\rho(B)-S_\rho(C)-S_\rho(ABC)\\
&=C^T(\rho)-I_\rho(A:B)-I_\rho(B:C)-I_\rho(C:A)\, , \label{eq:ve}
\end{align}
which is in accordance with the one considered by Kitaev and Preskill~\cite{PhysRevLett.96.110404}. 
Here, $S_\rho(A)$ stands for the von Neumann entropy of the RDM $\rho_A$ of region $A$ and $I_\rho(A:B)$ is the mutual information $I_\rho(A:B)=S_\rho(A)+S_\rho(B)-S_\rho(AB)$. For regions as given in Fig.~\ref{region}$(c)$ and assuming that there are no correlations between $A$ and $C$, i.e., $\rho_{AC}=\rho_A\ot\rho_C$, the above definition is consistent with the one by Levin and Wen~\cite{PhysRevLett.96.110405}. 
In the definition in Eq.~\eqref{deftopo}, the expectation of the total entropy is given by the Venn-diagram method, i.e., first summing up entropies of $AB$, $BC$, $CA$ and then subtracting entropies of overlapping regions.

On the other hand, from the information of the RDMs of local subsystems, we can estimate the entropy of the global state $\rho_{ABC}$ by using the maximum entropy method~\cite{PhysRev.106.620,PhysRev.108.171}. It is expected that the topological correlation in region $ABC$ can be measured by using $\trho^{(2)}_{ABC}$ as well as $\gamma$. 
Let us consider the 3rd-order irreducible correlation given by
\begin{align}\label{ic}
C^{(3)}(\rho_{ABC})= S_{\trho^{(2)}}({ABC})-S_\rho(ABC)\,.
\end{align}
By definition, the TEE and the irreducible correlation coincide if and only if 
\begin{align}\label{cortopo}
S_{\trho^{(2)}}({ABC})=&S_\rho(AB)+S_\rho(BC)+S_\rho(CA)\nonumber\\
&-S_\rho(A)-S_\rho(B)-S_\rho(C)\,.
\end{align}
While this equality does not hold in general, it is an interesting question whether Eq.~\eqref{cortopo} holds for ground states of gapped systems. 
In this paper, we will analytically show that the TEE and the irreducible correlation are equal if the spin model has zero correlation length. 

It is widely accepted that the ground state of a gapped system obeys an area law for the entanglement entropy of regions $A$ with smooth boundaries (see e.g.,~\cite{RevModPhys.82.277}), that is, 
\begin{equation}
S_\rho(A)=\alpha|\partial A|-n_{\partial A}\gamma+\cO(|\partial A|^{-1})\,.
\end{equation}
Here, $\alpha$ denotes a non-universal constant, $|\partial A|$ denotes the size of the boundary of region $A$ and $n_{\partial A}$ denotes the number of connected boundaries of $A$. 
If $|\partial A|$ is much larger than the correlation length, the contribution from local correlations $\cO(|\partial A|^{-1})$ can be ignored. 

In the following, we consider the case where the local contribution is {\it exactly} zero. This condition holds for fixed point wave functions in lattice models with zero-correlation length, such as quantum double models~\cite{Kitaev2003a} or Levin-Wen (string-net) models~\cite{PhysRevB.71.045110}. These models can describe a broad class of non-chiral topological orders. The crucial properties of these models can be summarized by the following two conditions:
\begin{itemize}
\item[(I)] If two regions $A$ and $B$ are separated, then the RDM is a product state $\rho_{AB}=\rho_A\ot\rho_B$, i.e., the mutual information $I_\rho(A:B)=0$. 
\item[(II)] If region $A$ and $C$ are indirectly connected through $B$ and $ABC$ has no holes, $\rho_{ABC}$ has zero conditional mutual information $I_\rho(A:C|B)\equiv I_\rho(A:BC)-I_\rho(A:B)=0$. 
\end{itemize}

A tripartite state that satisfies condition (II), i.e., $I_\rho(A:C|B)=0$, is referred to as a quantum Markov state (QMS) conditioned on $B$~\cite{ssaiff2004}. Such states have been widely studied in quantum information theory~\cite{ssaiff2004,RoQCMI,2014arXiv1410.0664F}, since they are the states  saturating strong subadditivity~\cite{qssapr}, that is, $S(\rho_{ABC}) = S_\rho(AB)+S_\rho(BC)-S_\rho(B)$.  Moreover, several applications as, for instance, in entanglement theory~\cite{SQAW1643788} and state redistribution are proposed~\cite{PhysRevLett.100.230501}.

Our main result is that if the ground state satisfies assumptions (I) and (II), the TEE is equivalent to the 3rd-order irreducible correlation. 
\bthm\label{thm1}
If assumptions (I) and (II) are satisfied, the equality
\begin{equation}\label{eq:thm1}
C^{(3)}(\rho_{ABC})=\gamma\,
\end{equation}
holds for all choices of regions $A$, $B$ and $C$ as depicted in Fig.~\ref{region}. 
\ethm
Note that this equivalence can further be generalized to more complicated regions with more holes or an annulus with more subregions. 

The 3rd-order irreducible correlation represents the distance of the tripartite state from the set of Gibbs states for all 2-local Hamiltonians. 
Therefore, this theorem implies that a non-zero value of the TEE is equivalent to that the entanglement Hamiltonian~\cite{PhysRevLett.101.010504} ${\tilde H}_{ABC}\equiv\log\rho_{ABC}$ on region $ABC$ cannot be a 2-local Hamiltonian. In other words, ${\tilde H}_{ABC}$ has to contain tripartite interactions acting on the whole region $ABC$.

Note that Theorem~\ref{thm1} together with Eq.~\eqref{eq:toc} and Eq.~\eqref{eq:ve} implies that the 2nd-order irreducible correlation $C^{(2)}(\rho)$ can be written as 
\begin{equation}
C^{(2)}(\rho)=I_\rho(A:B)+I_\rho(B:C)+I_\rho(C:A)\,.
\end{equation}
We finally mention that in Ref.~\cite{james215309} a measure for topological correlation that slightly differs from the irreducible correlation is proposed.  
The author claims that this quantity is also equivalent to the topological entanglement entropy if $A$ and $C$ in Fig.~\ref{region}$(c)$ satisfy $I_\rho(A:C)=0$, but our result indicates that it is only valid under additional assumptions. 

\subsection{Proof of the equivalence of the TEE and the irreducible correlation}
\begin{figure}[tb]
\begin{center}
\includegraphics[width=8cm]{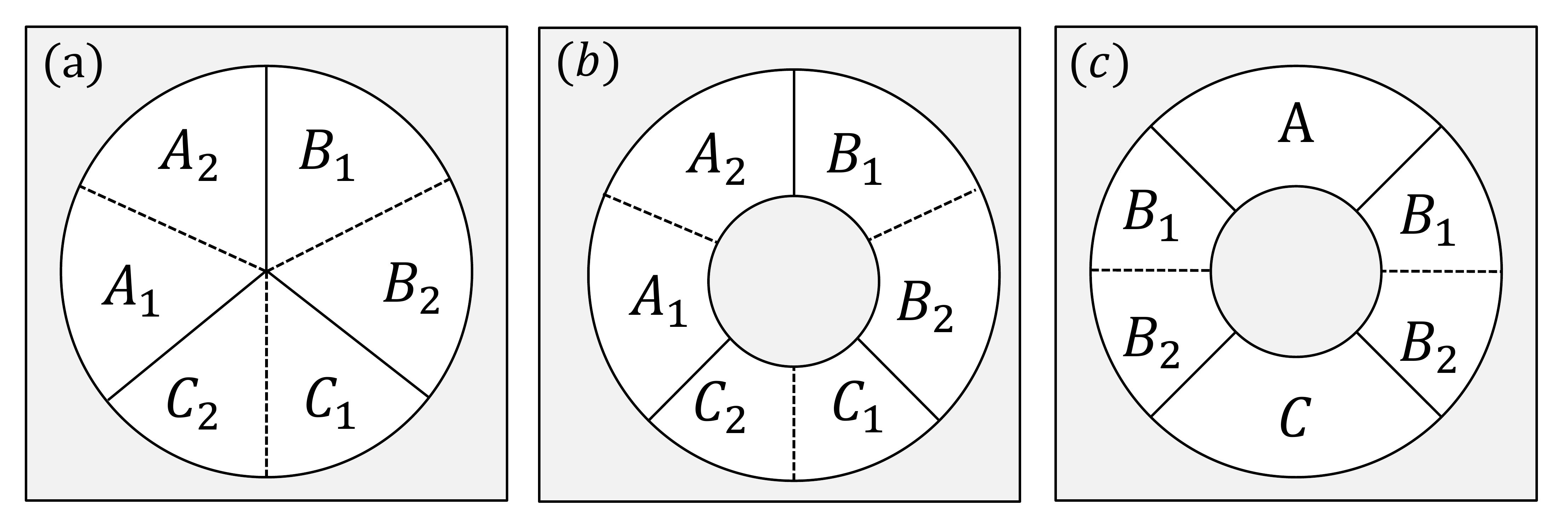}
\end{center}
\caption{A graphical illustration of how to divide the regions $A,B,C$ from Fig.~\ref{region} in the proof of Theorem~\ref{thm1}. Note that in any configuration $(a),(b),$ or $(c)$, the RDMs of three consecutive sub-regions are QMSs according to assumption (II). }
\label{regionsdiv}
\end{figure}

In order to prove Theorem 1, we use a strategy that explicitly exploits the structure of QMS. 
The sketch of the proof goes as follows. Let $A$, $B$ and $C$ be regions as illustrated in Fig.~\ref{region}. We then divide each region that connects two different regions into two halves (see Fig.~\ref{regionsdiv}). Then, each RDM on any three consecutive regions becomes a QMS according to property (II).  We then show that if properties (I) and (II) hold, we can merge these QMSs with overlapping local regions to obtain a global state. As the merged state is consistent to all local QMSs, it belongs to $R_\rho^2$,  and furthermore, it is a canonical candidate for the maximum entropy state since it is constructed using information of at most two regions. Indeed, we prove that the obtained global state is the maximum entropy state satisfying Eq.~\eqref{cortopo}, which  establishes Theorem 1. 

Let us show how to merge local QMSs to a global state. We use two basic properties of QMS shown in Ref.~\cite{ssaiff2004}. The first is that for any QMS $\rho_{ABC}$, there exists a recovery map $\Lambda_{B\to BC}$ such that 
\begin{equation}\label{recovery1}
\rho_{ABC}=(\id_A\ot\Lambda_{B\to BC})\rho_{AB}\,.
\end{equation}
The second property is a special form of the Koashi-Imoto decomposition~\cite{PhysRevA.66.022318} for QMSs. Namely, $\rho$ is a QMS if and only if there exists a decomposition of system $B$ into a direct sum $\cH_{B}=\bigoplus_i\cH_{B_{i}^L}\ot\cH_{B_{i}^R}$ 
such that~\cite{ssaiff2004} 
\begin{align}\label{markovdeco}
\rho_{ABC}=\bigoplus_ip_i\rho_{AB_i^L}\ot\rho_{B_i^RC}\,,
\end{align}
and $p_i$ is a probability distribution. We call Eq.~\eqref{markovdeco} a {\it Markov decomposition}.


We first consider the case of Fig.~\ref{regionsdiv} $(c)$. 
Since $\rho_{AB_1B_2}$ is a QMS conditioned on $B_1$, there exists a decomposition $\cH_{B_1}=\bigoplus_i\cH_{B_{1i}^L}\ot\cH_{B_{1i}^R}$ such that 
\begin{equation}\label{markovdeco2}
\rho_{AB_1B_2}=\bigoplus_ip_i\rho_{AB_{1i}^L}\ot\rho_{B_{1i}^RB_2}\,.
\end{equation}
Since $\rho_{B_1B_2C}$ is also a QMS, there exist a recovery map $\Lambda_{B_2\to B_2C}$ and a Markov decomposition such that 
\begin{align}
\rho_{B_1B_2C}&=(\id_{B_1}\ot\Lambda_{B_2\to B_2C})\rho_{B_1B_2}\label{recovery}\\
&=\bigoplus_jq_j\rho_{B_1B_{2j}^L}\ot\rho_{B_{2j}^RC}\,.\label{markovdeco3}
\end{align}
We then define the merged global state $\trho_{ABC}$ as
\begin{align}
{\tilde \rho}_{ABC}&\equiv(\id_{AB_1}\ot\Lambda_{B_2\to B_2C})\rho_{AB_1B_2}\,\label{2qmsdef}\\
&=\bigoplus_ip_i\rho_{AB_{1i}^L}\ot\Lambda_{B_2\to B_2C}(\rho_{B_{1i}^RB_2})\label{2qmsdeco}\,.
\end{align}
The second line follows because of Eq.~\eqref{markovdeco2}. Eq.~\eqref{2qmsdeco} represents a 
Markov decomposition of $\trho_{ABC}$. Hence, the state $\trho_{ABC}$ is a QMS conditioned on $B$.

We start by proving that ${\tilde \rho}_{ABC}$ can be decomposed as
\begin{align}
{\tilde \rho}_{ABC}&=\bigoplus_{i,j}p_iq_{j\vert i}\rho_{AB_{1i}^L}\ot\rho_{B_{1i}^RB_{2j}^L}\ot\rho_{B_{2j}^RC}\,,\label{2qms}
\end{align}
where $q_{j\vert i}\equiv\tr(\Pi_{B_{2j}}\rho_{B_{1i}^RB_2}\Pi_{B_{2j}})$ and $\Pi_{B_{2j}}$ denotes the orthogonal projector on $\cH_{B_{2j}^L}\ot\cH_{B_{2j}^R}$. 
In order to achieve this, we show that $\rho_{B_{1i}^RB_2}$ can be written as
\begin{equation}\label{decompbi}
\rho_{B_{1i}^RB_2}=\bigoplus_jq_{j\vert i}\rho_{B^R_{1i}B^L_{2j}}\ot\rho_{B_{2j}^R}\,,
\end{equation}
 where $\rho_{B_{1i}^RB_{2j}^L}$ is defined by
\begin{equation}
\rho_{B_{1i}^RB_{2j}^L}\equiv q_{j\vert i}^{-1}\tr_{B_{2j}^R}(\Pi_{B_{2j}}\rho_{B_{1i}^RB_2}\Pi_{B_{2j}}). 
\end{equation}
By definition it is clear that $q_{j|i}$ is a conditional probability distribution. Moreover, using the definition of $q_{j\vert i}$ and Eq.~\eqref{markovdeco3}, it is straightforward to check that $\sum_iq_{j\vert i}=q_j$. 

Let us consider the completely positive and trace-preserving (CPTP) map $P_2:\cS(\cH_{B_2})\to\cS(\cH_{B_2})$ defined by 
\begin{equation}
P_2(\xi_{B_2})=\bigoplus_j\tr_{B_{2j}^R}\left[\Pi_{B_{2j}}\xi_B\Pi_{B_{2j}}\right]\ot\rho_{B_{2j}^R}\,.
\end{equation}
Note that $({\id_{B_1}}\ot P_2)(\rho_{B_{1i}^L}\ot\rho_{B_{1i}^RB_2})=\rho_{B_{1i}^L}\ot\rho_{B_{1i}^RB_2}$ implies Eq.~\eqref{decompbi}. 
Eq.~\eqref{markovdeco3} yields that 
\begin{equation}\label{eq:Bdeco}
\rho_{B_1B_2}=\bigoplus_jq_j\rho_{B_1B_{2j}^L}\ot\rho_{B_{2j}^R}\,,
\end{equation}
from which follows that $({\id_{B_1}}\ot P_2)(\rho_B)=\rho_B$ holds. Owing to the invariance of $\rho_B$, we can  conclude 
\begin{align}
({\id_{B_1}}\ot P_2)&(\rho_{B_{1i}^L}\ot\rho_{B_{1i}^RB_2})\\&=({\id_{B_1}}\ot P_2)(p_i^{-1}\Pi_{B_{1i}^L}\rho_B\Pi_{B_{1i}^L})\\
&=p_i^{-1}\Pi_{B_{1i}^L}\left(({\id_{B_1}}\ot P_2)(\rho_B)\right)\Pi_{B_{1i}^L}\\
&=p_i^{-1}\Pi_{B_{1i}^L}\rho_B\Pi_{B_{1i}^L}=\rho_{B_{1i}^L}\ot\rho_{B_{1i}^RB_2}\,.
\end{align}
Thus, we have shown the decomposition  given by Eq.~\eqref{decompbi}.  Consequntly,  Eq.~\eqref{2qms} holds since the recovery map $\Lambda_{B_2\to B_2C}$ only acts on system $B_2^R$~\cite{ssaiff2004}.

\begin{figure}
\begin{center}
\vspace{-10mm}
\includegraphics[width=0.85\hsize]{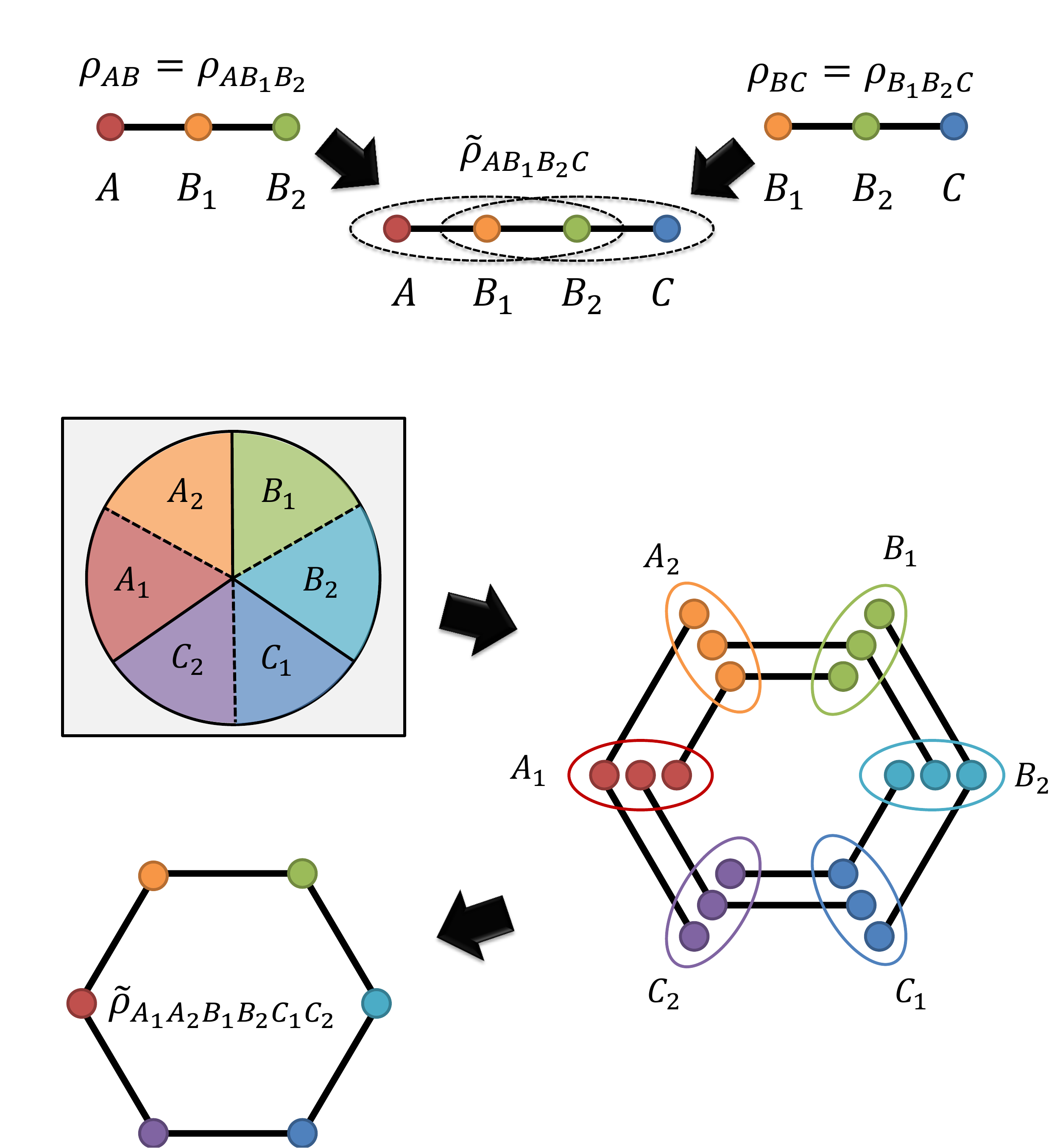}
\vspace{-5mm}
\end{center}
\caption{The upper illustration shows the merge of two local QMSs into one global QMS, which is consistent with the local original QMSs. The lower illustration shows the merge of six local QMSs in a cyclic way as used in the proof of Theorem~\ref{thm1} for regions as in Fig.~\ref{region} $(a)$ and $(b)$. }
\label{twoqms}
\vspace{-5mm}
\end{figure}

We now show that ${\tilde \rho}_{ABC}$ has the same 2-RDMs as $\rho_{ABC}$.  
${\tilde \rho}_{AB}=\rho_{AB}$ follows immediately from Eqs.~\eqref{markovdeco2}, \eqref{2qms} and \eqref{decompbi}. 
From the definition Eq.~\eqref{2qmsdef}, it turns out that  
\begin{equation}
\trho_{BC}=(\id_{B_1}\ot\Lambda_{B_2\to B_2C})\rho_{B_1B_2}=\rho_{BC}\,.
\end{equation}
The definition of $\trho_{ABC}$ and $I_\rho(A:B_2)=0$ implies that
\begin{align}
\trho_{AB_2C}&=\tr_{B_1}\left[(\id_{AB_1}\ot\Lambda_{B_2\to B_2C})\rho_{AB_1B_2}\right]\\
&=(\id_{A}\ot\Lambda_{B_2\to B_2C})\rho_{AB_2}\\&=(\id_{A}\ot\Lambda_{B_2\to B_2C})\rho_{A}\ot\rho_{B_2}\\
&=\rho_{A}\ot\Lambda_{B_2\to B_2C}(\rho_{B_2})=\rho_A\ot\rho_{B_2C}\,,
\end{align}
where we used in the third equality that $\rho_{AB_2}=\rho_A\ot\rho_{B_2}$ and $\Lambda_{B_2\to B_2C}(\rho_{B_2})=\rho_{B_2C}$ for the last equality.  
Therefore, $\trho_{AC}=\rho_A\ot\rho_C=\rho_{AC}$, which completes the proof that $\trho_{ABC}\in R_\rho^2$. 

It remains to show that $\trho_{ABC}$ is the maximum entropy state. 
Due to the strong subadditivity, for any state $\sigma\in R_\rho^2$, it holds that
\begin{align}
S_\sigma(ABC)&\leq S_\sigma(AB)+S_\sigma(BC)-S_\sigma(B)\\
&=S_\rho(AB)+S_\rho(BC)-S_\rho(B)&\\
&=S_\trho(ABC)\,
\end{align}
and thus, $\trho_{ABC}$ is the maximum entropy state in $R_\rho^2$.

Next, we consider the configuration encountered in Fig.~\ref{regionsdiv} $(a)$ and $(b)$, which is more involved as there exist six local QMSs (Fig.~\ref{twoqms}). 
For the following it is convenient to denote $A_1$ by $X_1$, $A_2$ by $X_2$, $B_1$ by $X_3$ and so on. Due to the periodicity, we consider the indices of $X_i$ modulo 6, i.e., $X_7=X_1$. 
For any neighboring three subregions ${X_{i-1}X_iX_{i+1}}$, the RDM $\rho_{X_{i-1}X_iX_{i+1}}$ is a QMS conditioned on $X_i$. Therefore, there exists a decomposition of $\cH_{X_i}=\bigoplus_{j_i}\cH_{X_i(j_i)}^L\ot\cH_{X_i(j_i)}^R$ such that $\rho_{X_{i-1}X_iX_{i+1}}$ can be written as
\begin{equation}
\rho_{X_{i-1}X_iX_{i+1}}=\bigoplus_{j_i}p_{j_i}\rho_{X_{i-1}X_{i(j_i)}^L}\ot\rho_{X_{i(j_i)}^RX_{i+1}}\,.
\end{equation} 
We denote the orthogonal projector on $\cH_{X_i(j_i)}^L\ot\cH_{X_i(j_i)}^R$ by $\Pi_{j_i}^{(i)}$. 
Our goal is to show that the maximum entropy state can be written as 
\begin{align}\label{QMring}
{\tilde \rho}_{ABC}=&\bigoplus_{i_1,...,i_6}p_1(i_1|i_6)p_2(i_2|i_1)\cdots p_6(i_6|i_5)\times\nonumber\\&\rho_{A_{1(i_1)}^RA_{2(i_2)}^L} \ot \rho_{A_{2(i_2)}^RB_{1(i_3)}^L}\ot\cdots\ot\rho_{C_{2(i_6)}^RA_{1(i_1)}^L}\, , 
\end{align}
where $p_j(i_j|i_{j-1})=\tr(\Pi_{i_j}^{(j)}\Pi_{i_{j-1}}^{(j-1)}\rho_{ABC})/\tr(\Pi_{i_{j-1}}^{(j-1)}\rho_{ABC})$.  
As long as it is clear from the arguments, we omit the lower index for the probabilities $p_j(i_j,i_j-1)$ and simply write $p(i_j,i_j-1)$.   

We show that under the assumptions (I) and (II), the cyclic products of conditional probabilities $p(i_1|i_6)p(i_2|i_1)\cdots p(i_6|i_5)$ form a probability distribution. 
The non-negativity is clear because each conditional probability is non-negative. The normalization condition can be shown by the following calculation
\begin{align}
&\sum_{i_1,...,i_6}p(i_1|i_6)p(i_2|i_1)\cdots p(i_6|i_5)\\
&=\sum_{i_2,...,i_6}\left(\frac{\sum_{i_1}p(i_6|i_1)p(i_2|i_1)p(i_1)}{p(i_6)}\right)p(i_3|i_2)\cdots p(i_6|i_5)\\
&=\sum_{i_2,...,i_6}\left(\sum_{i_1}\frac{p(i_6,i_1,i_2)}{p(i_6)}\right)p(i_3|i_2)\cdots p(i_6|i_5)\\
&=\sum_{i_2,...,i_6}\frac{p(i_6,i_2)}{p(i_6)}p(i_3|i_2)\cdots p(i_6|i_5)\\
&=\sum_{i_2,...,i_6}p(i_2)p(i_3|i_2)\cdots p(i_6|i_5)\\
&=\sum_{i_3,...,i_6}p(i_3)p(i_4|i_3)p(i_5|i_4)p(i_6|i_5)=\cdots=1\,.
\end{align}
The first equality follows from the Bayes rule $p(i|j)=p(j|i)p(i)/p(j)$. 
In the second equality, we used that $p(i_6,i_1,i_2)=p(i_1)p(i_6|i_1)p(i_2|i_1)$, which follows since $\rho_{C_2A_1A_2}$ is a QMS (i.e., assumption (II)) with the Markov decomposition
\begin{align}
\rho_{C_2A_1A_2}&=\bigoplus_{i_1}p(i_1)\rho_{C_{2}A_{1(i_1)}^L}\ot\rho_{A^R_{1(i_1)}A_2}\,.
\end{align}  
The fourth equality is due to $p(i_6,i_2)=p(i_6)p(i_2)$, which holds since $\rho_{C_2A_2}=\rho_{C_2}\ot\rho_{A_2}$ according to assumption (I). 

Now we are going to show that the state $\trho_{ABC}$ represented by Eq.~\eqref{QMring} is an element of $R_\rho^2$. 
Due to assumption (II), $\rho_{AB}$ is a QMS conditioned on $A_2$, $B_1$ and $A_2B_1$. Since a QMS is always a maximum entropy state,  $\rho_{AB}=\rho_{A_1(A_2B_1)B_2}$ has the same structure as the maximum entropy state in Eq.~\eqref{2qms}. Therefore, it can be decomposed as
\begin{align}
\rho_{AB}&=\bigoplus_{i_2,i_3}p(i_2)p(i_3|i_2)\rho_{A_1A^L_{2(i_2)}}\ot\rho_{A^R_{2(i_2)}B_{1(i_3)}^L}\ot\rho_{B^R_{1(i_3)}B_2}\,.
\end{align}
Similarly, it holds that 
\begin{align}
\rho_{BC}&=\bigoplus_{i_4,i_5}p(i_4)p(i_5|i_4)\rho_{B_1B^L_{2(i_4)}}\ot\rho_{B^R_{2(i_4)}C_{1(i_5)}^L}\ot\rho_{C^R_{1(i_5)}C_2}\,
\end{align}
and
\begin{align}\label{ACdeco}
\rho_{AC}&=\bigoplus_{i_6,i_1}p(i_6)p(i_1|i_6)\rho_{C_1C^L_{2(i_6)}}\ot\rho_{C^R_{2(i_6)}A_{1(i_1)}^L}\ot\rho_{A^R_{1(i_1)}A_2}\,.
\end{align}
Moreover, we can obtain a finer decomposition of each RDM by using decompositions~as Eq.~\eqref{decompbi}. For instance, by decomposing $C_1$ of  $\rho_{C_1C_{2(i_6)}^L}$ in Eq.~\eqref{ACdeco}, we obtain 
\begin{align}
\rho_{AC}=\bigoplus_{i_5,i_6,i_1}p(i_5)&p(i_6|i_5)p(i_1|i_6)\rho_{C^L_{1(i_5)}}\ot\rho_{C^R_{1(i_5)}C^L_{2(i_6)}}\nonumber\\
&\ot\rho_{C^R_{2(i_6)}A_{1(i_1)}^L}\ot\rho_{A^R_{1(i_1)}A_2}\,.
\end{align}

Without loss of generality, let us focus on the RDM $\trho_{AB}$, since the same arguments can be applied to system $BC$ and $CA$  due to the symmetry of the problem.  
We are going to show that $\tr_C\rho_{C^R_{2(i_6)}A_{1(i_1)}^L}$ and $\tr_C\rho_{B^R_{2(i_4)}C_{1(i_5)}^L}$ are independent of the indices $i_5$ and $i_6$ on $C$. These facts lead $\trho_{AB}=\rho_{AB}$. 
 Assumption (I) implies that $I_\rho(C_1^L:A_1)=0$. It further implies 
\begin{align}
\rho_{A_1C_1^L}&=\bigoplus_{i_5,i_6,i_1}p(i_5)p(i_6|i_5)p(i_1|i_6)\rho_{C_{1(i_5)}^L}\ot\rho^{i_6}_{A_{1(i_1)}^L}\ot\rho_{A^R_{1(i_1)}}\\
&=\bigoplus_{i_5,i_6}p(i_5)p(i_6|i_5)\rho_{C_{1(i_5)}^L}\ot\rho^{i_6}_{A_{1}}\\
&=\rho_{C_1^L}\ot\rho_{A_1}\,,
\end{align}
where $\rho^{i_6}_{A_{1(i_1)}^L}=\tr_{C^R_{2(i_6)}}\rho_{A_{1(i_1)}^LC^R_{2(i_6)}}$ and $\rho^{i_6}_{A_1}=\oplus_{i_1}p(i_1|i_6)\rho^{i_6}_{A_{1(i_1)}^L}\ot\rho_{A^R_{1(i_1)}}$. 
Therefore, $\rho^{i_6}_{A_{1}^L}$ must be independent of $i_6$.  Similarly, $I_\rho(B_2:C_2^R)=0$  implies 
\begin{align}
\rho_{B_2C_2^R}&=\bigoplus_{i_5,i_6}p(i_5)p(i_6|i_5)\rho_{B_2}^{i_5}\ot\rho_{C^R_{2(i_6)}}\\
&=\rho_{B_2}\ot\rho_{C_2^R}\,.
\end{align}
Therefore, $\rho_{B_2}^{i_5}$ must be independent of $i_5$. By tracing out 
 system $C$ of $\trho_{ABC}$, we obtain that 
\begin{align}
{\tilde \rho}_{AB}=&\bigoplus_{i_1,...,i_4}p(i_1)p(i_2|i_1)\cdots p(i_4|i_3)\rho_{A^L_{1(i_1)}}\ot\rho_{A^R_{1(i_1)}A^L_{2(i_2)}}\nonumber\\
&\qquad\quad\ot\rho_{A^R_{2(i_2)}B_{1(i_3)}^L}\ot\rho_{B^R_{1(i_3)}B^L_{2(i_4)}}\ot\rho_{B^R_{2(i_4)}}\\
&=\bigoplus_{i_2,...,i_4}p(i_2)p(i_3|i_2)p(i_4|i_3)\rho_{A_1A^L_{2(i_2)}}\ot\rho_{A^R_{2(i_2)}B_{1(i_3)}^L}\nonumber\\
&\qquad\quad\ot\rho_{B^R_{1(i_3)}B^L_{2(i_4)}}\ot\rho_{B^R_{2(i_4)}}\\
&=\bigoplus_{i_2,...,i_3}p(i_2)p(i_3|i_2)\rho_{A_1A^L_{2(i_2)}}\ot\rho_{A^R_{2(i_2)}B_{1(i_3)}^L}\nonumber\\
&\qquad\quad\ot\rho_{B^R_{1(i_3)}B_{2}}\\
&=\rho_{AB}\,.
\end{align} 
Note that in the first equality we used $\sum_{i_5,i_6}p(i_1|i_6)p(i_2|i_1)\cdots p(i_6|i_5)=p(i_1)p(i_2|i_1)\cdots p(i_4|i_3)$. The second equality follows from the Bayes rule
$p(i_1)p(i_2|i_1)=p(i_1|i_2)p(i_2)$ and the decomposition for $\rho_{A_1A^L_{2(i_2)}}$.  The third equality follows from the decomposition for $\rho_{B^R_{1(i_3)}B_2}$.

So far, we have shown that $\trho_{ABC}$ represented by Eq.~\eqref{QMring} is an element of $R_\rho^2$. It remains to prove that $\trho_{ABC}$ is the maximum entropy state. We rewrite $\trho_{ABC}$ in a more convenient form by defining new indices $a=(i_1,i_2), b=(i_3,i_4)$ and $c=(i_5,i_6)$ as
\begin{equation}\label{QMring2}
\trho_{ABC}=\bigoplus_{a,b,c}p(a|c)p(b|a)p(c|b)\rho_{A_a^RB_b^L}\ot\rho_{B_b^RC_c^L}\ot\rho_{C_c^RA_a^L}\,.
\end{equation}
Define the entanglement Hamiltonian $H_{ABC}=H_{AB}+H_{BC}+H_{CA}$, where 
\begin{align}
H_{AB}&=\sum_{a,b}\log[p(b|a)\rho_{A_a^RB_b^L}]\,,\\
H_{BC}&=\sum_{b,c}\log[p(c|b)\rho_{B_b^RC_c^L}]\,,\\
H_{AB}&=\sum_{a,c}\log[p(a|c)\rho_{C_c^RA_a^L}]\,.
\end{align}
By replacing zero eigenvalues in the logarithm in $H_{ABC}$ by a small positive constant $\epsilon$, we obtain the regularized 2-local Hamiltonian $H^\epsilon_{ABC}$.
It is easy to check that in the limit $\epsilon\to0$, $e^{H^\epsilon_{ABC}}$ converges  to $\trho_{ABC}$. According to \cite{weismaxent2015}, the maximum entropy state $\trho_{ABC}^{(2)}$ is the unique state in $R_\rho^2$ that can be represented as the limit of Gibbs  states of bounded 2-local Hamiltonians. Therefore,  $\trho_{ABC}$ is the maximum entropy state.   
\section{A Relation to Secret sharing of classical bits}\label{sec:ss} 
Using the equivalence of the TEE to the $3$rd-order irreducible correlation, we can now derive an operational interpretation of the TEE. Recall that if $C^{(3)}(\rho_{ABC})$ is nonzero, the global state in region $ABC$ cannot be uniquely determined from the marginals on $AB$, $BC$ or $AC$. 
A similar condition lies at the heart of secret sharing protocols. 
The goal of a $k$ out of $n$ secret sharing protocol is to share a classical (or quantum) secret among $n$ parties using a $n$-partite resource state such that groups of less than $k$ parties cannot read out the secret (see, e.g.,~\cite{PhysRevA.59.1829}). In particular, we consider a ramp scheme of secret sharing where we do not require the secret to be readable by any group of $k+1$ parties in contrast to the case of a threshold scheme.  In Ref.~\cite{2007quant.ph..1029Z,PhysRevLett.101.180505}, it is shown that for stabilizer states, the $k$th-order irreducible correlation represents the difference between the asymptotic bit rate that can be hidden from $k$ and from $k-1$ parties. 
We show that this also holds true in our setting for $n=3$ and $k=2$.

We consider a communication protocol for secret sharing and quantify the maximal asymptotic rate $R$ of secret bits that can be shared by using an infinite number of copies of a given resource state $\rho_{ABC}$. First, we fix the number of copies $N>0$. The sender chooses a secret $m$ in  $\cM_N=\{1,...,|\cM_N|\}$ and encodes it in a tripartite state according to a code-book $\{\rho^N_m\}$. The code states are given by states of the form $\rho^N_{m}=U_{m}\rho^{\ot N}_{ABC} U_{m}^\dagger$ satisfying $\rho^N_{m}\in R_{\rho^{\ot N}}^2$. 
The sender then distributes the tripartite state $\rho^{N}_m$ to three receivers to $A$, $B$ and $C$.  Since the bipartite RDMs of all code states are equal to the one of $\rho_{ABC}^{\ot N}$, the encoded secret $m$ can be read out only when all three receivers cooperate. In order to read the secret, the three receivers perform a global measurement described by a positive-operator valued measure (POVM)  $\{\Lambda_m^{(N)}\}$. The probability to falsely decode the message $m$  is $p^N(m)=\tr\{(\idty-\Lambda_m^{(N)})\rho^N_m\}$, and we denote the maximum error probability by $p^N_{\max}=\max_m p^N(m)$. 
 
We say that a secret sharing rate $r(\rho_{ABC})$ for $\rho_{ABC}$ is {\it achievable}, if for any $\delta,\epsilon>0$ and sufficiently large $N>0$, there exist an appropriate encoding method and a POVM such that $|\cM_N|=2^{N(r(\rho_{ABC})-\delta)}$ and $p^N_{\max}\leq\epsilon$. 
Owing to the Holevo-Schumacher-Westmoreland theorem~\cite{651037,PhysRevA.56.131}, the optimal secret sharing rate $R$ is obtained by
\begin{equation}\label{deftilde}
r(\rho_{ABC})=\lim_{N\to \infty}\frac{1}{N}\left[\max_{{\overline \rho^N}\in R_{\rho^{\ot N}}^2 }S\left({\overline \rho^N_{ABC}}\right)-S(\rho^{\ot N}_{ABC})\right]\,,
\end{equation}
where the maximum is taken over all uniformly distributed ensemble states ${\overline \rho^N_{ABC}}=\sum_{m}\frac{1}{M}U_{m}\rho^{\ot N}_{ABC} U^\dagger_{m}$ satisfying $U_{m}\rho^{\ot N}_{ABC} U^\dagger_{m}\in R_{\rho^{\ot N}}^2$ for all $i_N=1,...M$. 
The uniform distribution avoids a bias in the choice of the secret.

We then show the equivalence of the irreducible correlation to the optimal secret sharing rate: 
\bthm
For a tripartite state $\rho_{ABC}$ satisfying properties (I) and (II), the equality
\begin{equation}
r(\rho_{ABC})=C^{(3)}(\rho_{ABC})\,
\end{equation}
holds for all choices of regions depicted in Fig.~\ref{region}. 
\ethm 

To show the above equivalence, we generalize ideas from the proof for the bipartite~\cite{PhysRevA.74.042305} to the multipartite situation. 
Since the right hand side of Eq.~\eqref{deftilde} increases if the entropy of the state $\overline{\rho}^N$ increases, we need to find random unitary operations that conserve all bipartite RDMs and with an average state close to $\trho^{(2)\ot N}_{ABC}$.   
However, if the maximum entropy state is of form Eq.~\eqref{markovdeco} or Eq.~\eqref{QMring2}, that is, the state is a direct sum of product of local RDMs, we can find such a random unitary operation. 
\subsection{Proof of the equivalence between TEE and Optimal Secret Sharing Rate}
We first prove Theorem 2 for the case of Fig.~\ref{regionsdiv}$(c)$.
After that, we will generalize the proof to the other cases.
By assumption and the proof of Theorem 1, the maximal entropy state $\trho^{(2)}_{ABC}$ is a QMS conditioned on $B$ and can be decomposed as
\begin{equation}\label{thm2maxent}
\trho^{(2)}_{ABC}=\bigoplus_ip_i\rho_{AB_i^L}\ot\rho_{B_i^RC}\,.
\end{equation}

Let us consider the spectral decomposition of $\rho_{AB_i^L}$, that is, 
\begin{equation}\label{ABdeco2}
\rho_{AB_i^L}=\sum_{K_i}\lambda_{K_i}\Pi^{K_i}_{AB_i^L}\,,
\end{equation}
where $\Pi^{K_i}_{AB^L_i}$ is the projector on the eigensubspace corresponding to eigenvalue $\lambda_{K_i}$. More explicitly, $\Pi^{K_i}_{AB^L_i}$ can written as
\begin{equation}
\Pi^{K_i}_{AB_i^L}=\sum^{d_{K_i}}_{m_{K_i}=1}|K_i, m_{K_i}\rangle\langle K_i, m_{K_i}|_{AB_i^L}\,,
\end{equation}
where $|K_i,m_{K_i}\rangle$ are an orthonormal basis of the eigenspace of $\lambda_{K_i}$ and $d_{K_i}$ denotes the degeneracy. 
Then we expand the state $\rho_{ABC}$ by using eigenvectors of $\rho_{AB_i^L}$ to obtain
\begin{align}\label{thm2eq3}
\rho_{ABC}=\sum_{i,K_i,m_{K_i}}\sum_{j,L_j,n_{L_j}}|K_i,&m_{K_i}\rangle\langle L_j,n_{L_j}|_{AB^L}\nonumber\\&\ot w_{B^RC}^{i,K_i,m_{K_i},j,L_j,n_{L_j}}\,,
\end{align}
where $\cH_{B^L}=\bigoplus_i\cH_{B^L_i}$ and $\cH_{B^R}=\bigoplus_i\cH_{B^R_i}$. 

In the next step, we apply a random unitary $U_{AB^L}\in\cU$ of the form
\begin{equation}\label{runitary}
U_{AB^L}=\bigoplus_{i,K_i}U_{AB_i^L}^{K_i}\,,
\end{equation}
where  for every $i$ and $K_i$, $U_{AB_i^L}^{K_i}$ are drawn from an exact 1-design of the Haar measure on the eigenspace corresponding to the eigenvalue $\lambda$. 
Since all subspaces are finite-dimensional, the cardinality of $\cU$ is finite. 
According to Schur\rq{}s lemma, this random unitary operation transforms the state given by Eq.~\eqref{thm2eq3} to 
\begin{equation}\label{eq:barrho}
{\bar \rho}_{ABC}=\bigoplus_{i,K_i,m_{K_i}}\Pi^{K_i}_{AB_i^L}\ot w_{B_i^RC}^{i,K_i,m_{K_i}}\,,
\end{equation}
where $w_{B_i^RC}^{i,K_i,m_{K_i}}=w_{B^RC}^{i,K_i,m_{K_i},i,K_i,m_{K_i}}$.  
Since $\trho^{(2)}$ and ${\bar \rho}$ have same 2-RDMs as $\rho$, we obtain
\begin{align}
\trho_{AB^L}&=\bigoplus_{i,K_i}p_i\lambda_{K_i}\Pi^{K_i}_{AB_i^L}\\
&={\bar \rho}_{AB^L}\\
&=\bigoplus_{i,K_i}\tr\left(\sum_{m_{K_i}} w_{B_i^RC}^{i,K_i,m_{K_i}}\right)\Pi^{K_i}_{AB_i^L}\,.
\end{align}
 Thus, it holds that
\begin{equation}
\tr\left(\sum_{m_{K_i}} w_{B_i^RC}^{i,K_i,m_{K_i}}\right)=p_i\lambda_{K_i}\,.
\end{equation}
We denote the normalized operator $\frac{1}{p_i\lambda_{K_i}}\sum_{m_{K_i}} w_{B_i^RC}^{i,K_i,m_{K_i}}$ by $\rho_{B_i^RC}^{K_i}$. 
Note that $\rho_{B_i^RC}=\sum_{K_i}q_{K_i}\rho_{B_i^RC}^{K_i}$, where $q_{K_i}=\lambda_{K_i}d_{K_i}$, but the states in $\{\rho_{B_i^RC}^{K_i}\}$ are not necessarily orthogonal to each other. 
Then, ${\bar \rho}_{ABC}$ can be written as
\begin{equation}\label{eq:bardecom}
{\bar \rho}_{ABC}=\bigoplus_{i,K_i}p_i\lambda_{K_i}\Pi^{K_i}_{AB_i^L}\ot \rho_{B_i^RC}^{K_i}\,.
\end{equation}
The difference between $\bar\rho$ and $\trho^{(2)}$ is that $\bar\rho$ has additional correlations between $AB_i^L$ and $B_i^RC$ via the index $K_i$. 

Summarizing the above calculations, we obtain an ensemble of states $\{\frac{1}{|\cU|}, U_i\rho_{ABC} U_i^\dagger\in R_\rho^k\}$ where the entropy of the averaged state ${\bar \rho}_{ABC}$ is given by
\begin{align}\label{entbar}
S({\bar \rho}_{ABC})=&H(\{p_i\})+\sum_ip_iH(\{q_{K_i}\})\\&+\sum_{i,K_i}p_iq_{K_i}\left(\log d_{K_i}+S(\rho_{B_i^RC}^{K_i})\right)\,.
\end{align}
From  Eqs.~\eqref{thm2maxent} and \eqref{ABdeco2}, the entropy of $\trho^{(2)}$ is given by
\begin{align}\label{entmax}
S(\trho^{(2)}_{ABC})=&H(\{p_i\})+\sum_ip_iH(\{q_{K_i}\})\\&+\sum_{i,K_i}p_iq_{K_i}\log d_{K_i}+\sum_ip_iS(\rho_{B_i^RC})\,.
\end{align}
By taking the difference between Eqs.~\eqref{entbar} and \eqref{entmax}, the 3rd-order irreducible correlation of ${\bar \rho}_{ABC}$ can be bounded by
\begin{align}
C^{(3)}({\bar\rho}_{ABC})&=S(\trho^{(2)}_{ABC})-S({\bar \rho}_{ABC})\\
&=\sum_ip_i\left[S(\rho_{B_i^RC})-\sum_{K_i}q_{K_i}S(\rho_{B_i^RC}^{K_i})\right]\\
&=\sum_ip_i\left[S(\sum_{K_i}q_{K_i}\rho^{K_i}_{B_i^RC})-\sum_{K_i}q_{K_i}S(\rho_{B_i^RC}^{K_i})\right]\\
&\leq\sum_ip_iH(\{q_{K_i}\})\leq\max_{i}\log D_{i}\\&\leq\log D\,,
\end{align}
where $D_i$ and $D$ denote the number of different eigenvalues of $\rho_{AB_i^L}$ and $\rho_{AB^L}$, respectively. 
If we consider $N$ copies of $\rho_{ABC}$, $D$ grows only polynomially in $N$, whereas the total dimension of the Hilbert space grows exponentially. 
If the dimension of the Hilbert space $\cH_A\ot\cH_{B^L}$ is denoted by $d_{AB^L}$, the number of eigenvalues $D^N$ of the $N$-copy state  $\rho^{\ot N}_{ABC}$ is bounded by~\cite{Wilde201304},
\begin{equation}
D^N\leq \log(N+1)^{d_{AB^L}}\,.
\end{equation}

Given the expression for the rate 
\begin{equation}
r(\rho_{ABC})=\lim_{N\to \infty}\frac{1}{N}\left[\max_{{\overline \rho^N}\in R_{\rho^{\ot N}}^2 }S\left({\overline \rho^N_{ABC}}\right)-S(\rho^{\ot N}_{ABC})\right]\,,
\end{equation}
 and using that the irreducible correlation is additive, we obtain 
\begin{align}
r(\rho_{ABC})&=\lim_{N\to \infty}\frac{1}{N}\left[\max_{{\overline \rho^N}\in R_{\rho^{\ot N}}^2 }S\left({\overline \rho^N_{ABC}}\right)-S(\rho^{\ot N}_{ABC})\right]\label{RR}\\
&=\lim_{N\to \infty}\frac{1}{N}\left[\max_{{\overline \rho^N}\in R_{\rho^{\ot N}}^2 }S\left({\overline \rho^N_{ABC}}\right)-S(\trho^{(2)\ot N})\right]\\
&\qquad+S(\trho^{(2)}_{ABC})-S(\rho^{\ot N}_{ABC})\\
&=C^{(3)}(\rho_{ABC})-\lim_{N\to \infty}\frac{1}{N}\max_{{\overline \rho^N}\in R_{\rho^{\ot N}}^2 }C^{(3)}\left({\overline \rho^N_{ABC}}\right)\\
&\geq C^{(3)}(\rho_{ABC})-\lim_{N\to \infty}\frac{1}{N}\log(N+1)^{d_{AB^L}}\\
&=C^{(3)}(\rho_{ABC})\,.
\end{align}
This establishes a lower bound on the optimal rate $R$ by $C^{(3)}$. However, 
the upper bound $r(\rho_{ABC})\leq C^{(3)}(\rho_{ABC})$ follows directly from Eq.~\eqref{RR}, and the definition of $C^{(3)}(\rho_{ABC})$. This completes the proof.

In the case of Fig.1$(b)$ and $(c)$, i.e., the maximum entropy state can be written as Eq.~\eqref{QMring2}, 
we iteratively perform  random unitary operations as discussed in the previous case to systems $AB$ and $AC$. Let us rewrite $\trho_{ABC}^{(2)}$ as 
\begin{equation}
\hspace{-0.5mm}\trho^{(2)}_{ABC}=\bigoplus_{a,b,c}p(a,b)p(c|a,b)\rho_{A_a^RB_b^L}\ot\rho_{B_b^RC_c^L}\ot\rho_{C_c^RA_a^L}\,,\hspace{-2mm}
\end{equation}
where $p(c|a,b)=p(c|a)p(c|b)/p(c)$. We then introduce the spectral decomposition 
$\rho_{A_a^RB_b^L}=\sum_{K_{ab}}\lambda_{K_{ab}}\Pi_{A_a^RB_b^L}^{K_{ab}}$.  Let us define a set of unitaries $\{U_{A^RB^L}\}$ in the same way as in the previous case. Consequently, the averaged state becomes
\begin{align}
{\bar\rho}_{ABC}=\bigoplus_{a,b, K_{ab}}p(a,b)\lambda_{K_{ab}}\Pi^{K_{ab}}_{A_a^RB_b^L}\ot\rho^{K_{ab}}_{A_a^LB_b^RC}\,
\end{align}
for some state $\rho^{K_{ab}}_{A_a^LB_b^RC}$. We further introduce the spectral decomposition $\rho_{C_c^RA_a^L}=\sum_{L_{ac}}\mu_{L_{ac}}\Pi_{C_c^RA_a^L}^{L_{ac}}$ and a set of unitaries $\{U_{C^RA^L}\}$ similar to $\{U_{A^RB^L}\}$. After performing the second average over the unitaries $\{U_{C^RA^L}\}$, the state can be written as
\begin{align}
{\bar{\bar\rho}}_{ABC}=\bigoplus_{a,b,c, K_{ab},L_{ac}}&p(a,b)p(c|a,b)\lambda_{K_{ab}}\mu_{L_{ac}}\times\nonumber\\
&\Pi^{K_{ab}}_{A_a^RB_b^L}\ot\Pi^{L_{ac}}_{C_c^RA_a^L}\ot\rho^{K_{ab},L_{ac}}_{B_b^RC_c^L}\,.
\end{align}
Since the remaining correlation in ${\bar{\bar\rho}}_{ABC}$ is also bounded by the logarithm of the number  $K_{ab}, L_{ac}$ of different eigenvalues, we can use the same argument as in the case of Fig.1$(c)$. Therefore, Theorem 2 holds for all situations presented in Fig.1. 
\section{Approximately vanishing correlations}\label{sec:Approx}
In general, assumption (I) and (II) are not perfectly satisfied and there are small local correlations between separated regions. These correlations only vanish in the thermodynamic limit. 
We are interested in whether the TEE and the irreducible correlation are close if the correlations are sufficiently small i.e., each region is sufficiently larger than the correlation length. 
Unfortunately, our proofs cannot be generalized straightforwardly to this situation and we cannot answer this question completely.  
However, by introducing a \lq\lq{}smoothed\rq\rq{} version of the irreducible correlation, 
we can show that at least the Levin-Wen type TEE and the \lq\lq{}smoothed\rq\rq{} irreducible correlation are close. 

To discuss finite deviation due to the local correlations, 
it may be useful to define a set of multipartite states where their $k$-RDMs are $\delta$-similar to $\rho$ as 
 \begin{equation}\label{def:mR}
R_\rho^{k,\delta}\equiv\left\{\sigma\in\cS(\cH^{n}) \mid \forall S_k \;\rm{ s.t. } \;|S_k|=k, \,\|\sigma_{S_k}-\rho_{S_k}\|_\tr\leq\delta\right\}\,.
\end{equation}
Then, we define the $\delta$-variation of the irreducible correlation as
\begin{equation}\label{def:mirr}
C_\delta^{(k)}(\rho)\equiv S(\trho^{(k-1),\delta})-S(\trho^{(k),\delta})\,,
\end{equation}
where $\trho^{(k),\delta}$ is the state having maximum entropy among all states in the closed convex set $R_\rho^{k,\delta}$.

In order to generalize our results to finite correlation lengths, we have to relax the condition that two far apart regions have exactly zero correlation to the case that the correlation is arbitrary small. 
While $I_\rho(A:B|C) \approx 0$ does not guarantee that $\rho_{ABC}$ is close to a state with a Markov decomposition in Eq.~\eqref{markovdeco}~\cite{RoQCMI}, it has been discovered~\cite{2014arXiv1410.0664F} that $I_\rho(A:C|B)\approx0$ implies that there exists a recovery map  $\Lambda_{B\to BC}$ such that
\begin{equation}
\rho_{ABC}\approx(\id_A\ot\Lambda_{B\to BC})\rho_{AB}\,.
\end{equation}
By using this result, it is possible to extend our argument for the case of Fig.~\ref{regionsdiv} $(c)$ if the assumptions are satisfied with small error $\delta$. 
In fact, one can obtain that 
\begin{equation}\label{eq:difa1}
|C^{(3)}_\delta(\rho)-I_\rho(A:C|B)|\leq f(\delta)\,
\end{equation}
for some function $f(\delta)$ which goes 0 in the limit $\delta\to0$ (see the appendix for details). 
Hence, if $\delta$ is sufficiently small, the $\delta$-variation of the irreducible correlation and the Levin-Wen type TEE are close. However, $|C_\delta^{(k)}(\rho)-C^{(k)}(\rho)|$ is not necessarily to be small due to the discontinuity of the irreducible correlation~\cite{weisdiscont2012}. 
Also, the proof of the Kitaev-Preskill type TEE  is more involved since the maximum entropy state cannot be written in terms of the recovery maps of local QMSs. Moreover, a way to extend the relation to the optimal secret rate is unclear since the proof fully relies on the Markov decomposition.  

\section{Conclusion and open problems}\label{sec:Conclusion}
We have presented an information-theoretical approach to analyze the TEE of states with zero correlation lengths. 
In particular, we have established the equivalence between the TEE, the irreducible correlation and the optimal secret sharing rate. 
Via the irreducible correlation we obtain an interpretation of the TEE in both
Kitaev-Preskill's and Levin-Wen's approaches as the distance of the ground state to the set of Gibbs states corresponding to Hamiltonians with only bipartite interactions.
This means that a non-zero TEE implies that the reduced
state on ABC (see Fig. 1) contains genuine tripartite correlations in the sense that the reduced state cannot be
approximated by a Gibbs state of a Hamiltonian with
only bipartite interactions. 
Moreover, the equivalence to the optimal secrete sharing rate provides an intuitive operational meaning to the TEE as the amount of information that can be encoded in topologically non-trivial global regions without being detectable by access to any partial  (i.e., topologically trivial) regions.

Although we only show our results for exactly vanishing correlation lengths, we expect that they also hold approximately if the local correlations are vanishing approximately, i.e., (I) and (II) are not satisfied perfectly. Unfortunately, our techniques based on QMS do not allow us to generalize our result straightforwardly in this direction suggesting that new technical tools are required. 
Thanks to recent breakthroughs in the study of quantum states with small conditional mutual information~\cite{2014arXiv1410.0664F,2015arXiv150407251S}, 
we can show that the Levin-Wen type TEE is close to a smoothed version of the irreducible correlation. 
But due to the lack of continuity of the irreducible correlation, this does not suffice to prove that in general the TEE and the irreducible correlation are close. 
Moreover, it is not clear how to extend our result to the Kitaev-Preskill type TEE (i.e., Fig.~\ref{region} (a)), since the maximum entropy state cannot be represented via recovery maps.  

Our results motivate to investigate further the relation between the irreducible correlation and characteristic properties of topological orders such as long-range entanglement and locally indistinguishable ground states~\cite{PhysRevLett.97.050401, PhysRevLett.111.080503}. Furthermore, it is known that the TEE is related to the total quantum dimension~\cite{PhysRevLett.96.110404} of the corresponding anyonic model. It would thus be interesting to derive such a relation from a more operational approach using the interpretation of the TEE as the optimal secrete sharing rate. To do so, the Wilson loop operators, which are non-local operators related to the quantum dimensions of anyonic charges, might be utilized as global encoding operators.

Besides, another interesting question is whether the operational interpretation of the irreducible correlation to the optimal secret sharing rate extends to general quantum multipartite states. If the equivalence holds, it provides a useful formula to obtain an operational decomposition of the $n$-partite total correlation via the maximum entropy principle.

\section*{Acknowledgment}
This work is supported by ALPS, 
the Project for Developing Innovation Systems
of MEXT, Japan, and JSPS KAKENHI
(Grant No. 26330006 and No.15H01677).
We also gratefully acknowledge the ELC project (Grant-in-Aid for
Scientific Research on Innovative Areas MEXT KAKENHI (Grant No.
24106009)) for encouraging the research presented in this paper. 
FF acknowledges support from Japan Society for the Promotion of Science (JSPS) by KAKENHI grant No. 24-02793. 
\appendix
\section{Proof of Eq.~\eqref{eq:difa1}}
In realistic models of topological ordered phases, our assumptions $(I)$ and $(II)$ hold only approximately due to local interactions. 
In the following we restrict our consideration for the region given in Fig.~\ref{regionsdiv} $(c)$. Namely, we assume that
\begin{equation}\label{asu1}
I_\rho(A:B_2C)\leq\epsilon
\end{equation}
and 
\begin{align}
I_\rho(A:B_2|B_1)&\leq\epsilon,\label{asu2}\\
I_\rho(B_1:C|B_2)&\leq\epsilon\label{asu22}
\end{align}
hold, and we choose $\epsilon$ as an upper bound for all these quantities. 

By using Pinsker\rq{}s inequality, the first assumption~\eqref{asu1} implies that
\begin{equation}
\|\rho_{AB_2C}-\rho_A\ot\rho_{B_2C}\|_\tr\leq2\sqrt{\epsilon}\,.
\end{equation}
By using the monotonicity of the trace norm, this also implies that $\|\rho_{AC}-\rho_A\ot\rho_C\|_\tr\leq2\sqrt{\epsilon}$.

The following theorem about the recovery maps recently proven in Ref.~\cite{2015arXiv150407251S} is crucial in the proof.
\bthm~\cite{2015arXiv150407251S}\label{thmA}
For any state $\rho_{BC}$ on $\cH_B\ot\cH_C$ there exists a CPTP map (recovery map) $\Lambda_{B\to BC}$ such that for any state $\rho_{ABC}$ satisfying $\tr_A\rho_{ABC}=\rho_{BC}$ 
\begin{equation}
I_\rho(A:C|B)\geq-2\log_2F\left(\rho_{ABC}, (\id_A\ot\Lambda_{B\to BC})\rho_{AB}\right)\,,
\end{equation}
where $F(\rho,\sigma)=\tr[\sqrt{\sqrt{\rho}\sigma\sqrt{\rho}}]$ is the fidelity between $\rho$ and $\sigma$.
\ethm
Note that the fidelity satisfies $\|\rho-\sigma\|_\tr\leq2\sqrt{1-F^2}$. 
Therefore the assumptions~\eqref{asu2} and~\eqref{asu22} imply that there exists CPTP maps $\Lambda_{B_1\to AB_1}$ and $\Lambda_{B_2\to B_2C}$ such that 
\begin{align}
\left\|\rho_{AB_1B_2}-(\Lambda_{B_1\to AB_1}\ot \id_{B_2})\rho_{B_1B_2}\right\|_\tr\leq2\sqrt{1-2^{-\epsilon}}\,,\label{approx1}\\
\left\|\rho_{B_1B_2C}-(\id_{B_1}\ot\Lambda_{B_2\to B_2C})\rho_{B_1B_2}\right\|_\tr\leq2\sqrt{1-2^{-\epsilon}}\label{approx2}\,.
\end{align}

Similar to the proof of Theorem~\ref{thm1} in the main text, we define a global state $\trho_{ABC}$ by
\begin{equation}\label{defapg}
\trho_{ABC}\equiv(\id_{AB_1}\ot\Lambda_{B_2\to B_2C})\rho_{AB_1B_2}\,.
\end{equation}
The 2-RDMs of this global state is close to the original state $\rho$. By tracing out system $A$ in Eq.~\eqref{defapg}, we obtain that
\begin{equation}
\trho_{B_1B_2C}=(\id_{B_1}\ot\Lambda_{B_2\to B_2C})\rho_{B_1B_2}\,,
\end{equation}
and therefore we have
\begin{equation}
\left\|\rho_{B_1B_2C}-\trho_{B_1B_2C}\right\|_\tr\leq2\sqrt{1-2^{-\epsilon}}\,
\end{equation}
because of Eq.~\eqref{approx2}. Combining Eq.~\eqref{approx1}, Eq.~\eqref{approx2} and Eq.~\eqref{defapg} yields
\begin{align*}
\trho_{AB_1B_2}&\approx\tr_C(\id_{AB_1}\ot\Lambda_{B_2\to B_2C})(\Lambda_{B_1\to AB_1}\ot \id_{B_2})\rho_{B_1B_2}\\
&=\tr_C(\Lambda_{B_1\to AB_1}\ot \id_{B_2C})(\id_{B_1}\ot\Lambda_{B_2\to B_2C})\rho_{B_1B_2}\\
&\approx\tr_C(\Lambda_{B_1\to AB_1}\ot \id_{B_2C})\rho_{B_1B_2C}\\
&=(\Lambda_{B_1\to AB_1}\ot \id_{B_2})\rho_{B_1B_2}\\
&\approx\rho_{AB_1B_2}\,.
\end{align*}
The precise calculation is performed by using the triangle inequality and the monotonicity of the trace norm. As a result, we obtain
\begin{equation}\label{apxAB}
\left\|\rho_{AB_1B_2}-\trho_{AB_1B_2}\right\|_\tr\leq6\sqrt{1-2^{-\epsilon}}\,.
\end{equation}
Finally, since $\rho_{AB_2}\approx\rho_A\ot\rho_{B_2}$, taking the partial trace over $B$ yields
\begin{align*}
\trho_{AC}&=\tr_{B_2}\left[(\id_{A}\ot\Lambda_{B_2\to B_2C})\rho_{AB_2}\right]\\
&\approx\tr_{B_2}\left[(\id_{A}\ot\Lambda_{B_2\to B_2C})(\rho_A\ot\rho_{B_2})\right]\\
&=\tr_{B_2}(\rho_A\ot\rho_{B_2C})\\
&=\rho_A\ot\rho_C\\
&\approx\rho_{AC}\,.
\end{align*}
In the third line, we used Theorem~\ref{thmA} applied to the tripartite state $\rho_A\ot\rho_{B_2C}$. 
More precisely, we obtain that
\begin{equation}
\|\rho_{AC}-\trho_{AC}\|_\tr\leq4\sqrt{\epsilon}\,.
\end{equation}
Since $4\sqrt{\epsilon}\leq6\sqrt{1-2^{-\epsilon}}$, we conclude that 
\begin{equation}
\trho_{ABC}\in R_\rho^{2,\delta}\,
\end{equation}
for $\delta=6\sqrt{1-2^{-\epsilon}}$. 

So far we have constructed a global state $\trho_{ABC}$ with 2-RDMs similar to $\rho_{ABC}$. 
Although this is not the maximum entropy state in $R_\rho^{2,\delta}$, we can obtain good bounds of $C^{(3)}_\delta$ from this state.  
To do so, we use the fact that a state that is approximately recoverable has small conditional mutual information~\cite{2014arXiv1410.0664F}. 
Eq.~\eqref{apxAB} implies that
\begin{equation}
\|\trho_{ABC}-(\id_{AB_1}\ot\Lambda_{B_2\to B_2C})\trho_{AB}\|_\tr\leq\delta\,
\end{equation}
by definition of $\trho_{ABC}$. From Eq.~$(10)$ in  Ref.~\cite{2014arXiv1410.0664F} , we obtain that
\begin{align}
I_\trho(A:C|B)
\leq7\log_2d_A\sqrt{\delta}
\end{align}
for sufficiently small $\delta$. Therefore, by using strong subadditivity,
Therefore, we find that for any state (including the maximum entropy state)  $\sigma_{ABC}\in R_\rho^{2,\delta}$, 
\begin{align}
S_{\sigma}(ABC)&\leq S_\sigma(AB)+S_\sigma(BC)-S_\sigma(B)\\
&\leq S_{\trho}(AB)+S_{\trho}(BC)-S_{\trho}(B)\nonumber\\
&\quad+2\delta\log{d_Ad_B^2d_C}+3\eta(2\delta)\\
&\leq S_{\trho}(ABC) +2\delta\log{d_Ad_B^2d_C}\nonumber\\
&\quad+3\eta(2\delta)+7\sqrt{\delta}\log_2d_A\\
&\equiv S_{\trho}(ABC)+\frac{1}{2}f(\delta)\,,\label{eq:deltairr}
\end{align}
where $\eta(x)=-x\log(x)$. The first line follows by the strong subadditivity. In the second line, we used the triangle inequality to obtain $\Vert \sigma_{AB}-\trho_{AB}\Vert\leq2\delta$ and then used the Fannes inequality. The third line follows by Theorem~\ref{thmA}. 
Note that $\lim_{\delta\to 0}f(\delta)=0$.
In conclusion, it holds that 
\begin{align}
C^{(3)}_\delta(\rho)&\leq S_{\trho}(ABC)+\frac{1}{2}f(\delta)-S_\rho(ABC)\nonumber\\
&\quad+2\delta\log{d_Ad_B^2d_C}+3\eta(2\delta)\label{eq:a41}\\
&\leq I_\rho(A:C|B)+f(\delta)\,,
\end{align}
where we again used the Fannes inequality.

Since the maximum entropy state in $R_\rho^{2,\delta}$ has entropy larger or equal to $\trho_{ABC}$,  inequality~\eqref{eq:deltairr} also implies that
\begin{align}
C^{(3)}_\delta(\rho)&\geq S(\trho_{ABC})-S(\rho_{ABC})\\
&\geq S_\rho(AB)+S_\rho(BC)-S_\rho(B)\nonumber\\
&\quad-\frac{1}{2}f(\delta)-S(\rho_{ABC})\\
&\geq I_\rho(A:C|B)-\frac{1}{2}f(\delta)\,.
\end{align}
Hence, we conclude that 
\begin{equation}
|C^{(3)}_\delta(\rho)-I_\rho(A:C|B)|\leq f(\delta)\,
\end{equation}
holds for sufficiently small $\delta$.

\end{document}